# Transferring Axial Molecular Chirality Through a Sequence of On-Surface Reactions

*Néstor Merino-Díez[1,2,3,‡], Mohammed S. G. Mohammed[1,3,‡], Jesús Castro-Esteban[4,‡], Luciano Colazzo[1,3,†], Alejandro Berdonces-Layunta[1,3], James Lawrence[1,3], J. Ignacio Pascual[2,5], Dimas G. de Oteyza[1,3,5,\* and Diego Peña[4,\*]*

1. Donostia International Physics Center (DIPC); 20018 San Sebastián, Spain.

2. CIC nanoGUNE, nanoscience cooperative research center; 20018 San Sebastián, Spain.

3. Centro de Física de Materiales-Material Physics Center (CFM-PCM), 20018 San Sebastián, Spain.

4. Ikerbasque, Basque Foundation for Science; 20018 San Sebastián, Spain.

5. CiQUS, Centro Singular de Investigación en Química Biolóxica e Materiais Moleculares; 15705 Santiago de Compostela, Spain.

**Corresponding Author**

\* d_g_oteyza@ehu.eus, \* diego.pena@usc.es

**Present Addresses**

† Center for Quantum Nanoscience, Institute for Basic Science (IBS), Seoul 03760, Republic of Korea; Department of Physics, Ewha Womans University, Seoul 03760, Republic of Korea.

**Author Contributions**

**ABSTRACT:** Fine management of chiral processes on solid surfaces has progressed over the years, yet still faces the need for the controlled and selective production of advanced chiral materials. Here, we report on the use of enantiomerically enriched molecular building blocks to demonstrate the transmission of their intrinsic chirality along a sequence of on-surface reactions. Triggered by thermal annealing, the on-surface reactions induced in this experiment involve firstly the coupling of the chiral reactants into chiral polymers and subsequently their transformation into planar prochiral graphene nanoribbons. Our study reveals that the axial chirality of the reactant is not only transferred to the polymers, but also to the planar chirality of the graphene nanoribbon end products. Such chirality transfer consequently allows, starting from adequate enantioenriched reactants, for the controlled production of chiral and prochiral organic nanoarchitectures with pre-defined handedness.

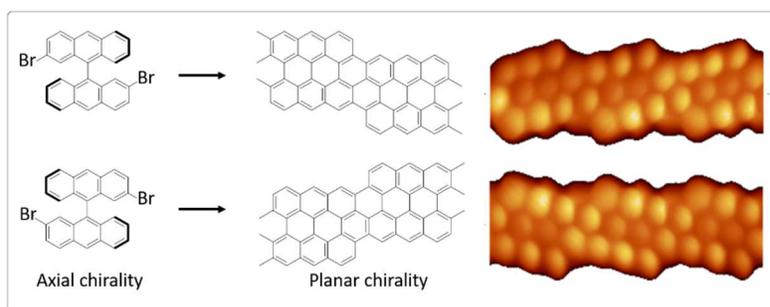

**TOC**

The relevance of molecular chirality is well-known in several branches of science.[1] Particularly focusing on chemistry, enantioselective synthesis, which aims at obtaining a majority of one enantiomer of a chiral product, emerged long ago as a top-class division in academic, industrial and pharmaceutical chemistry.[2] In this context, solid surfaces appeared as a good candidate for the development of heterogeneous enantioselective catalysts.[3] This established chirality as a topic of interest within the field of surface science, profiting from the application of new analytical techniques. One such example is scanning probe microscopy, which allows identification of the adsorbate's chirality readily at the single molecule level,[4] in contrast to the most conventional method, which requires the analyte´s crystallization (in turn needing relatively large quantities) and subsequent X-ray diffraction analysis.

Early on-surface experiments addressing chirality in ultra-high vacuum (UHV) conditions reported on the formation of chiral nanostructures (molecular domains, clusters and/or single molecules) after deposition of molecular racemates on different surfaces, analyzed by means of low-energy electron diffraction,[5,6] scanning tunneling microscopy (STM),[4,7–9] and atomic force microscopy.[10,11] These forerunning works evidenced the presence of additional intriguing aspects of chirality on surfaces such as, for example, the chirality arising from the mirror-symmetry breaking upon surface physisorption of non-chiral (or *achiral*) molecules.[12,13] Nowadays, such issues are extensively reviewed.[14–18]

However, in spite of the booming interest in surface-supported organic chemistry, typically termed as "on-surface synthesis",[19] the transmission of chirality through on-surface reactions is still scarcely explored. Pioneering work from De Schryver and coworkers reported on the conservation of adsorption-induced chirality from self-assembled domains of diacetylene molecules to homochiral polymeric lattices,[20] as recently did Chi and coworkers using alkylated benzenes.[21] Focusing on intrinsic molecular chirality (rather than adsorption-induced), particularly noteworthy is the use of helical aromatic molecules (so-called *helicenes*) to study the diastereoselective formation of helical dimers[22–24] or the transmission from the helical chirality of the molecular precursors to the planar chirality (prochirality) of nanographene adsorbates through a sequence of single-molecule reactions.[25] However, no studies have yet reported e.g. the transfer of axial chirality, nor a chirality analysis across polymerization reactions of chiral reactants.

Inspired in earlier works,[26–28] in 2016 we showed that the on-surface polymerization of 2,2'-dibromo-9,9'-bianthracene (DBBA) led to bianthryl polymers that could subsequently be transformed into prochiral (3,1)-graphene nanoribbons (GNRs) by cyclodehydrogenation.[29] The reactant and polymer intermediate display axial chirality, while the end product is prochiral, and this multistep on-surface synthesis process was proved successful on different achiral surfaces, namely on Cu(111), Ag(111) and Au(111).[29] Although products of both prochiralities were observed, they never mixed in a single GNR.[30] Bearing in mind that the enantiomers of DBBA are atropisomers which exhibit axial chirality ($R$ or $S$ enantiomers, Fig. 1), this presumably relates to the steric hindrance between the radicals generated from different enantiomers as they approach, which prevents their Ullmann coupling while allowing the polymerization of monomers that share the same chirality (sketched in Fig. S1). This raises the question whether each enantiomer could selectively lead to a different chiral GNR adsorbate. At this point, it is important to remark that a random adsorption process of prochiral nanostructures like (3,1)-GNRs on achiral surfaces cannot favor a particular handedness. However, the use of an enantiomerically enriched molecular precursor, in combination with chirality transfer, may allow for an unbalanced generation of one particular enantiomer of the prochiral GNRs. The ability of selecting the handedness of prochiral GNRs on a surface would entitle us to direct with precision the synthesis of advanced chiral-selective graphene structures,[31] or the interfacing with other molecular species in hybrid systems.[32,33]

With this idea in mind, we separated both DBBA enantiomers from the racemic mixture by HPLC (see Fig. S2), verifying the different optical activity of each enantiomer (indicated in Fig. 1, bottom labels). We were thus able to isolate enantiomerically enriched samples of (+) and (-)-DBBA (98:2 enantiomeric ratio) which were independently deposited on the surface. Analysis by STM allowed us to establish the absolute configuration of the enantiomers ($S$ for (+)-DBBA; $R$ for (-)-DBBA) and to study how the inherent handedness of the precursors evolves through the different reactions performed on the substrate. The suggested chirality transfer process is sketched in Fig. 1.

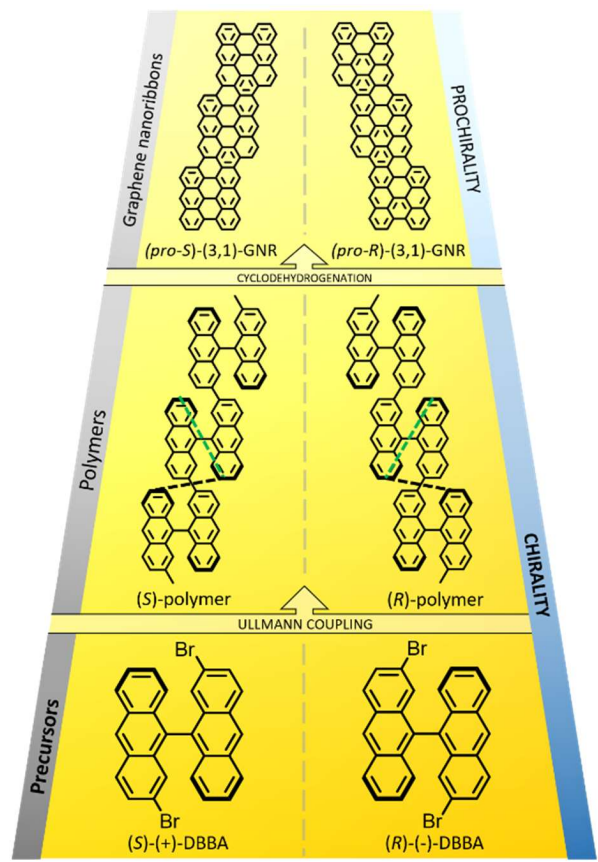

**Fig. 1.** Scheme of the different reactions accounted in this work. (Bottom) Both enantiomers of the non-reacted chiral GNR precursors. (Middle) Chiral anthracene-based polymers. (Top) Prochiral graphene nanoribbons. The non-equivalent distances ($d_1$ and $d_2$) between up-pointing anthracene ends in the polymeric structure are marked in green and black, respectively.

After precursor deposition, we thermally induced Ullmann coupling,[34] which caused the enantiomers to fuse into polymers.[29] Figure 2a shows a representative image of the system after the polymerization of enantioenriched (*S*)-(+)-DBBA, where polymers appear aggregated into islands. The reactant's chirality is transferred to the polymer, which displays a non-planar structure with alternatingly tilted anthracene units due to the steric hindrance between hydrogen atoms. Imaged with a scanning tunneling microscope, the result is a zigzagging chain of round features corresponding to the up-pointing ends of the anthracene units.[29,30] The slightly asymmetric intramolecular distances between those features ($d_1$ and $d_2$ in Fig. 2b) can be discerned and associated to each polymeric enantiomer (see Fig. 1), allowing for the identification of their absolute configuration. In addition, in the absence of undesired species and/or defects, the handedness of these structures can be identified or cross-checked also by analyzing their longitudinal ends, whose oblique orientation depends on the polymer's chirality (Fig. 2c). Figures S3 and S4 show a comparison between the intramolecular distances and longitudinal ends' oblique orientation of both chiral polymers *R* and *S*. Overall, the unambiguous chirality determination allows us to quantify the enantiomeric ratio, revealing an excess of 95.7 % for (*S*)-polymers (Fig. 2d).

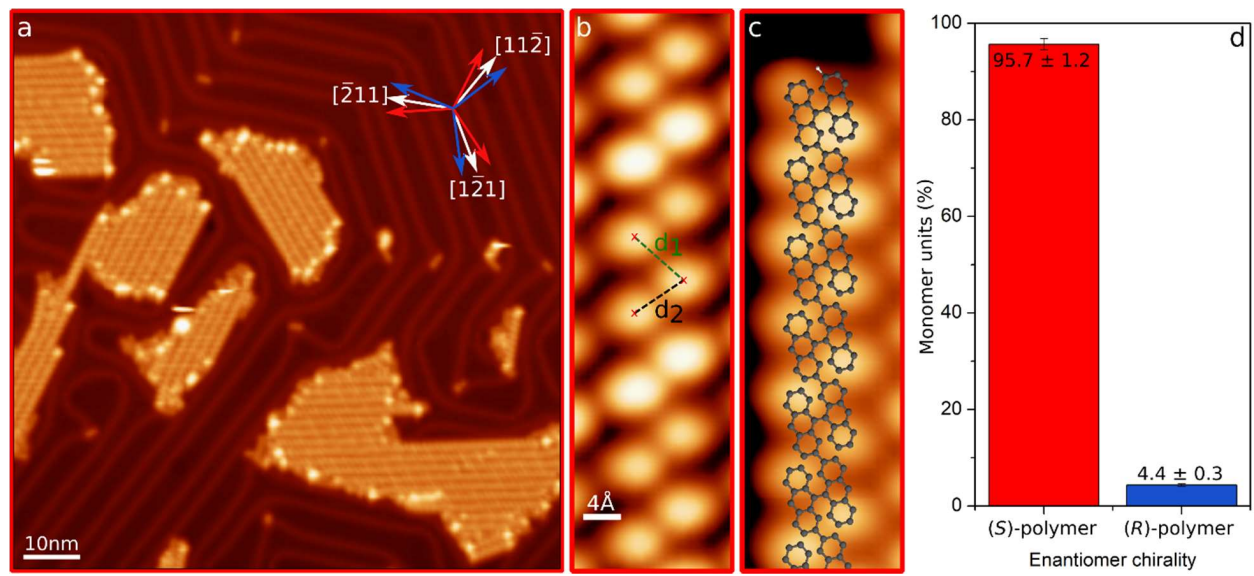

**Fig. 2. a)** Representative STM overview image ($U_s$ = 1.0 V, $I_t$ = 32 pA) of the sample after polymerization of enantioenriched (*S*)-(+)-DBBA reactants. The inset indicates the three favored growth orientations of each chiral polymer (red and blue arrows) with respect to Au(111) crystallographic directions (white arrows). **b,c)** STM images of chiral (*S*)-polymer ($U_s$ = 0.5 V, $I_t$ = 10/50 pA). Dashed lines in (b) represent the non-equivalent intramolecular distances *d1* and *d2*. A superimposed structure model is included in (c). **d)** Percentage of monomer's chirality found in polymers after polymerization of the enantioenriched (*S*)-(+)-DBBA precursor.

As marked with the arrows in Fig. 2a, substrate-adsorbate interactions steer the adsorption of each enantiomeric polymer into three orientations with respect to the Au(111) crystallographic directions, holding a mirror-symmetry relation (whose mirror plane is aligned with the Au(111) high-symmetry directions) that demonstrates their chiral nature. Interestingly, this growth preference is altered for single polymers or islands of small size (Fig. S5), or even for larger islands when polymers of opposite chirality aggregate together. Independently of island dimensions, the surface reconstruction is lifted below the polymers and the soliton lines modified so as to surround the islands. This effect is attributed to the strong interaction of halogen atoms with the surface, as previously reported for similar systems.[35–37] Although not visible by STM except along the sides of the islands, it is known from XPS that halogens are still present on the surface,[29] presumably in between the non-planar polymers[38] (which thus "hide" them from the scanning probe). They are thus responsible, as also found with other hydrocarbon polymers,[39,40] for the attractive interpolymer interactions that drive the island formation.

Higher temperature annealing triggers the cyclodehydrogenation (CDH) of the polymers, transforming them into planar prochiral graphene nanoribbons.[29,30] For the sake of simplicity, we maintain the same nomenclature *R/S* when referring to each configuration. Figure 3a shows a representative overview of the sample after the planarization of the enantioenriched (*S*)-polymer sample. As for the preceding polymers, the adsorbate-substrate interactions drive an epitaxial alignment along three well-defined orientations for ribbons of each prochirality.[30] By functionalizing the metallic STM tip with a CO molecule we can gain distinct resolution of the planar adsorbate's bonding structures[41,42] (Fig. 3b-d, Fig. S6 includes images of both configurations). Such unambiguous determination of the prochirality again allows us to quantify the enantiomeric ratio of the GNRs, revealing an excess of 93.4 % for prochiral (*S*)-(3,1)-GNRs ('*pro-S'*). In a control experiment making use of enantioenriched *R*-reactants, we observe a numerically comparable enantiomeric excess, this time of prochiral (*R*)-(3,1)-GNRs ('*pro-R'*) (Fig. S7).

At this point, we compare and analyze the evolution of the enantiomeric ratios at each stage across the complex multistep reaction process. The HPLC analysis of reactants rendered a S:R enantiomeric ratio of 98:2 (Fig. S2), while STM analysis (see the methods section) rendered 95.7 ± 1.2 : 4.4 ± 0.3 for the chiral polymers and 93.4 ± 2.1 : 6.6 ± 0.6 for prochiral GNRs. Although the enantiomeric excess seems to be slightly decreasing across subsequent stages, the changes remain close to the error margins, making this decrease nearly meaningless. Therefore, our results unambiguously confirm that the reactant´s chirality is transferred across the various substrate-supported reactions in spite of the increasingly high activation temperatures. The barrier for monomer racemization is thus exceedingly high to be crossed neither in the gas phase during sublimation at 435 K nor on the surface held at 415 K for polymerization. After polymerization, the presumably increased racemization barrier (requiring a concerted change of all monomers along the polymer) is not accessible either at the cyclodehydrogenatio temperature of 625 K. Although a racemization of the GNRs could also come from GNRs or polymers 'fliping' over on the surface, such process will intuitively show an energy barrier not far from that of desorption, which on the other hand is observed to be unsubstantial for the system and the temperatures employed.

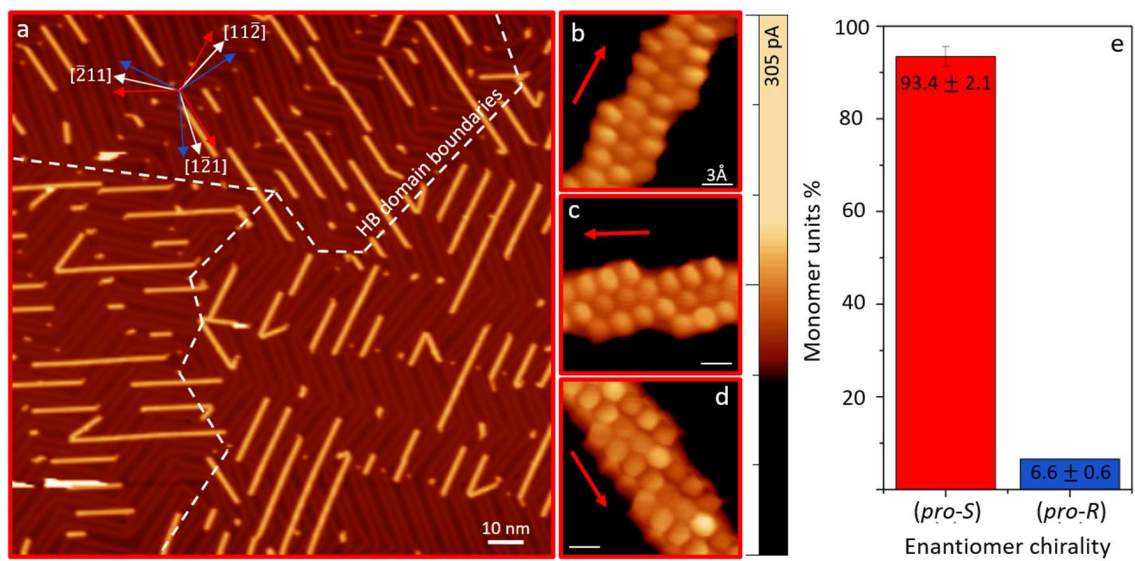

**Figure 3.** a) Representative STM overview image ($U_s$ = 0.5 V, $I_t$ = 40 pA) of the GNR sample after cyclodehydrogenation of the primarily (*S*)-polymer sample shown in Fig. 2. The inset indicates the three growth orientations of each enantiomeric polymer (red and blue arrows) with respect to the high symmetry Au(111) directions (white arrows). The boundaries between different azimuthal domains of the Au(111) herringbone reconstruction are marked with white dashed lines, evidencing a preferential GNR orientation for each domain. b-d) Constant-height current maps ($U_s$ = 2 mV) with a CO-terminated tip showing (*pro-S*)-GNRs in their three growth orientations on Au(111). e) Enantiomeric distribution of the monomers forming the GNRs.

Finally, Fig. 3a also reveals other important details, such as (i) the recovery of the Au(111) herringbone reconstruction throughout the whole surface, (ii) the unidirectional alignment of chiral GNRs within each herringbone reconstruction domain and (iii) a well-defined favored interspacing between parallel ribbons. A key point related to these findings is the desorption of Br when the cyclodehydrogenation sets in.[43] As the strongly interacting halogens leave the surface, the reconstruction reappears. Besides, also the halogen-mediated intermolecular interactions vanish, no longer driving the adsorbate´s agglomeration into islands. The templating effect of the recovered reconstruction on the now independently diffusing GNRs determines the ribbon's alignment and well-defined interspacing. On the one hand, as previously observed with many other aromatic adsorbates,[44–46] also for GNRs[47] the slightly higher electron potential in the reconstruction´s face-centered-cubic (*fcc*) regions with respect to the hexagonal-closed-packing (*hcp*) regions[48] cause a favored adsorption on the former. On the other hand, the ribbon's epitaxy on Au(111) displays a preferred adsorption orientation at 14 degrees from the directions followed by the fcc trenches ($[11\bar{2}]$ and symmetry-related directions).[30] Thus, one of the three epitaxially equivalent orientations for GNRs of each prochirality is favored by maximizing its adsorption length on *fcc* sections as compared to the other two, which would display more soliton crossing points and *hcp* adsorption regions. Given the large enantiomeric excess of one chirality, a clearly dominating GNR orientation is observed on each herringbone domain (Fig. 3a, Fig. S7). This also explains the observation of preferential GNR interspacings. We measure a preferred distance of 6.8 ± 0.3 nm (or multiples thereof) independently of GNRs prochirality or herringbone domain (Fig. S8). Taking into account the herringbone reconstruction´s periodicity of ≈ 6.34 nm, and the 14 degrees deviation of the ribbons with respect to the herringbone soliton lines, the reconstruction's periodicity perpendicular to the GNRs' orientation is ≈ 6.53 nm, in close agreement to the observed GNR spacing.

In conclusion, we report the transfer of the intrinsic axial chirality of enantiomerically enriched molecular precursors to the resulting prochiral graphene nanoribbons through a multistep on-surface synthesis process that also involves chiral polymers as intermediates. The reactants present sufficiently high energy barriers to prevent conformational changes that could result in their racemization across the various surface-assisted chemical reactions. As a result, an unbalanced generation of prochiral (3,1)-GNR with one particular adsorption configuration has been achieved. The use of chirality transfer allows for the controlled production of chiral organic nanoarchitectures and their combination with other molecular species in complex systems. Such control paves the way for the study of physicochemical phenomena directly related to their chiral nature, like heterogeneous enantioselective catalysis and/or optical activity; and therefore for the development of forefront devices such as spin filters[49] or circularly polarized light detectors.[50]

## EXPERIMENTAL METHODS

**Sample preparation.** 2,2′-Dibromo-9,9′-bianthracene (DBBA) was synthesized following the procedure previously described.[25] For the preparation of the different samples, the enantiomers of DBBA were independently evaporated at 435 K from a home-made Knudsen cell oriented towards the monocrystalline Au(111) surface substrate for deposition. Atomically cleaned Au(111) surface were obtained by standard sputtering (0.8 kV, $Ar^+$) and annealing cycles (705K). Thermally-induced on-surface reactions were performed by radiative heating, at 415K (525K) for Ullmann coupling (cyclodehydrogenation).

**STM imaging.** Sample analysis was performed in a commercial Scienta-Omicron low-temperature STM under ultrahigh vacuum (UHV) conditions with pressure values below the $10^{-10}$ mbar range and a base temperature of 4.3 K. All STM images were processed with the WSxM software.[51] The chirality determination from the STM images was performed as described in the text, and the provided statistics are the result of counting the number of monomer units (each accounting for one unit cell in polymers and GNRs) observed in polymers or GNRs of each handedness. For larger scale images, once the chirality of a polymer or GNR is determined, the number of monomers is estimated from the polymer/GNR length divided by the respective unit cell size. The reported errors stem from the square root of the total monomer counts for each species and chirality.

## ASSOCIATED CONTENT

**Supporting Information**. 3D models comparing the steric hindrance between homochiral and heterochiral reactants, enantiomeric partition of reactants, additional STM images comparing different chiralities in polymers and GNRs, control experiment with enantioenriched $S_a$-reactants, large-scale STM images showing GNR distribution. This material is available free of charge via the Internet at http://pubs.acs.org.

ACKNOWLEDGMENT

We acknowledge funding from the European Union's Horizon 2020 programme (Grant Agreement Nos. 635919 ("SURFINK") and 863098 ("SPRING")), from the Spanish MINECO (Grant No. MAT2016-78293-C6), Xunta de Galicia (Centro singular de investigación de Galicia, accreditation 2016-2019, ED431G/09), and the Fondo Europeo de Desarrollo Regional (FEDER).

# Supporting Information:

# Transferring axial molecular chirality through a sequence of on-surface reactions

*Néstor Merino-Díez[1,2,3,‡], Mohammed S. G. Mohammed[1,3,‡], Jesús Castro-Esteban[5,‡], Luciano Colazzo[1,3], Alejandro Berdonces-Layunta[1,3], James Lawrence[1,3], J. Ignacio Pascual[2], Dimas G. de Oteyza[1,3,4],\* and Diego Peña[5,]\**

1. Donostia International Physics Center (DIPC); 20018 San Sebastián, Spain.
2. CIC nanoGUNE, nanoscience cooperative research center; 20018 San Sebastián, Spain.
3. Centro de Física de Materiales-Material Physics Center (CFM-PCM), 20018 San Sebastián, Spain.
4. Ikerbasque, Basque Foundation for Science; 20018 San Sebastián, Spain.
5. CiQUS, Centro Singular de Investigación en Química Biolóxica e Materiais Moleculares; 15705 Santiago de Compostela, Spain.

## Table of Contents

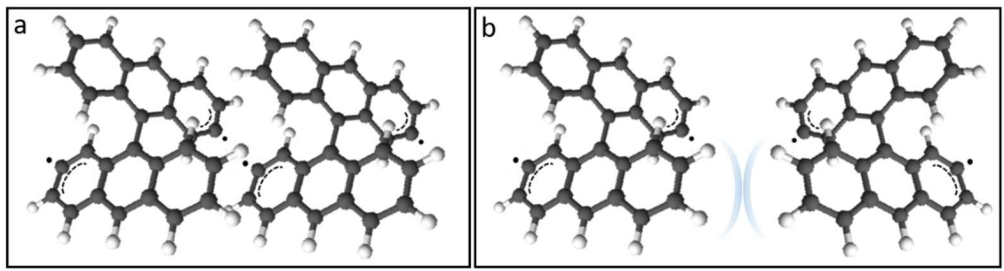

**Fig. S1.** 3D models illustrating the steric hindrance exerted between dehalogenated monomers of (a) the same and (b) different chirality. The latter displays a strong steric hindrance that hinders polymerization of monomers with opposite chirality.

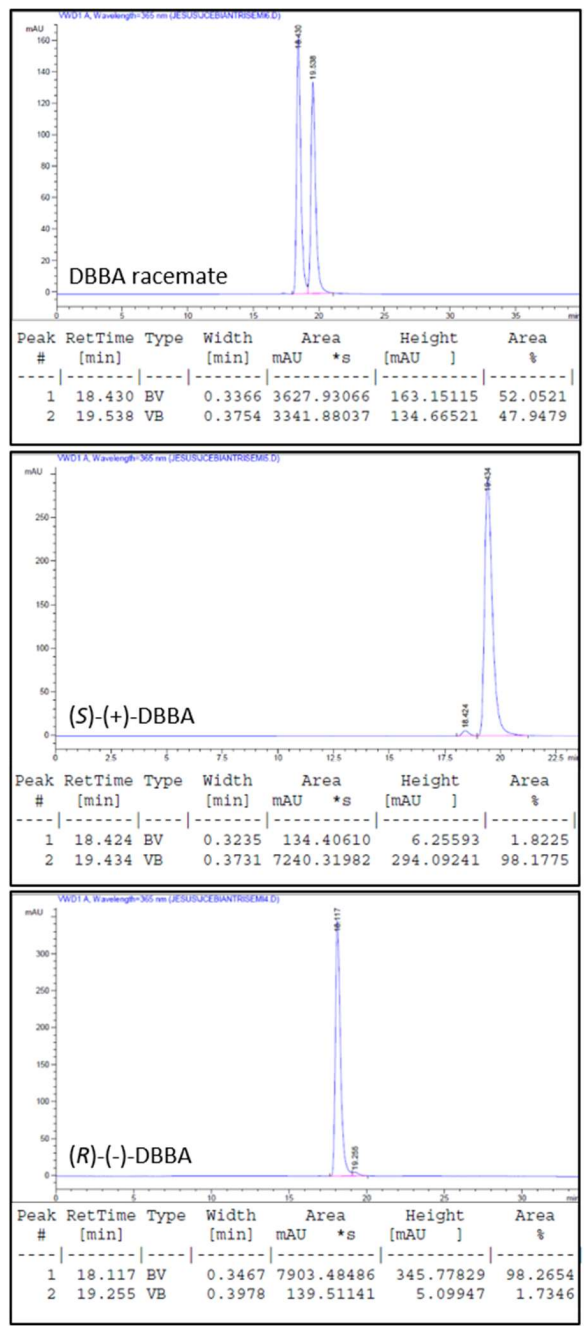

**Fig. S2.** Chiral HPLC profiles for (top) DBBA racemate, and enantiomerically enriched (middle) (S)-(+)- and (bottom) (R)-(-)-DBBA precursors.

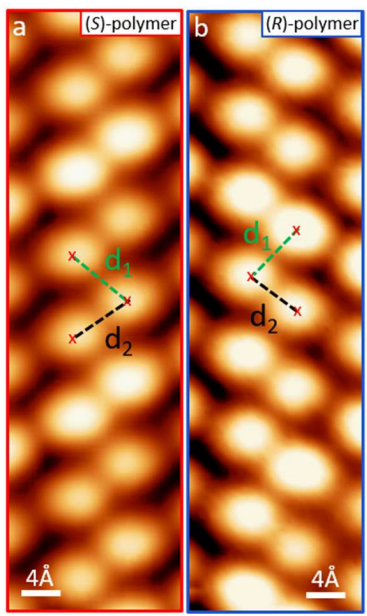

**Fig. S3.** Identification of the polymer's chirality from their non-equivalent intramolecular distances between the uppointing ends of the tilted anthracene moieties. Examples (a) (S)-polymer and (b) (R)-polymer ($V_s$ = 0.5 V, $I_t$ = 10-50 pA).

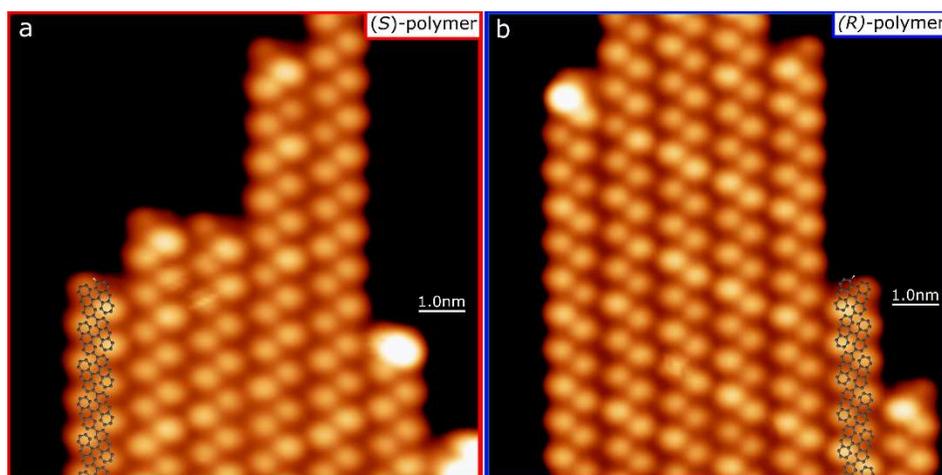

**Fig. S4.** Identification of the polymer's chirality from their longitudinal end oblique orientation. Examples ($V_s$ = 0.5 V, $I_t$ = 50 pA) of (a) (S)-polymer and (b) (R)-polymer with superimposed models where, for an easier visualization, just terminal hydrogen atoms (in white) are represented.

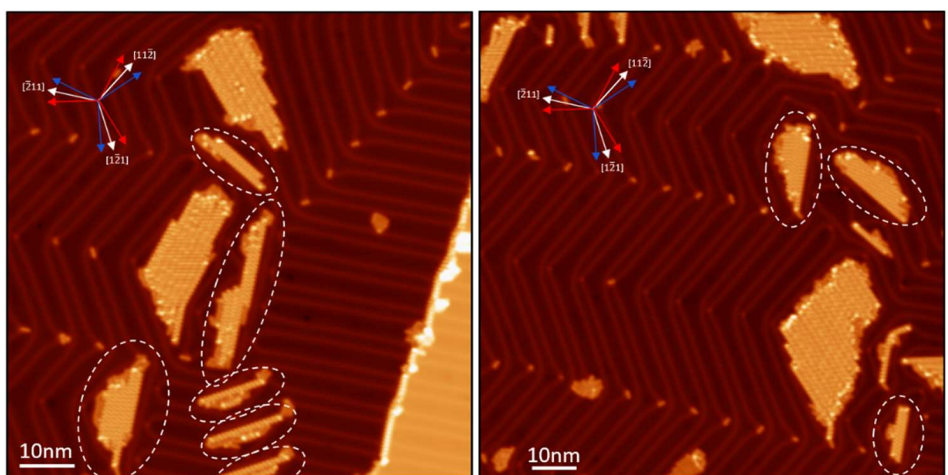

**Fig. S5.** STM images ($V_s$ = 1.0 V, $I_t$ = 32 pA) showing examples of (S)-polymers islands out of main growth orientations (circled in white dashed lines).

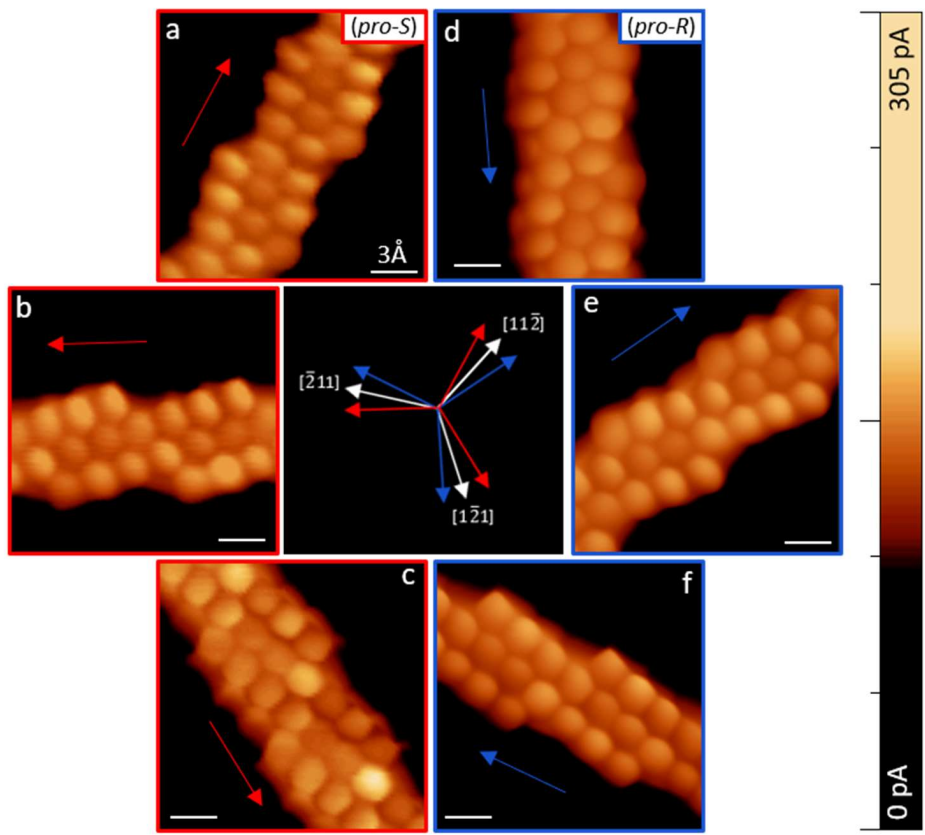

**Fig. S6.** Constant-height current maps ($V_s$ = 2 mV) with a CO-terminated tip showing (a-c) (*pro-S*)-GNRs and (d-f) (*pro-R*)-GNRs in their three growth orientations on Au(111), as indicated in the central inset.

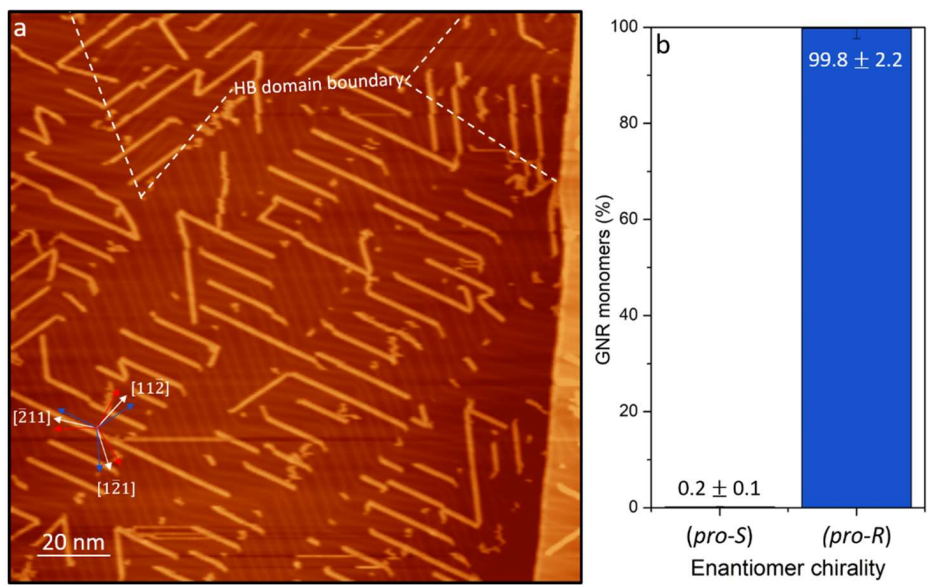

**Fig. S7.** a) Representative STM overview image ($V_s$ = 0.1 V, $I_t$ = 50 pA) of the sample after depositing enantioenriched (*R*)-(-)-DBBA precursor and inducing its polymerization and cyclohydrogenation. White dashed lines represent the boundaries of the herringbone reconstruction domains that determine the preferred orientations of GNRs. Inset indicates the preferred adsoption orientations of prochiral GNRs with respect to Au(111) . b) Enantiomeric distribution of the monomers forming the GNRs after polymerization and cyclohydrogenation of enantioenriched (*R*)-(-)-DBBA precursors.

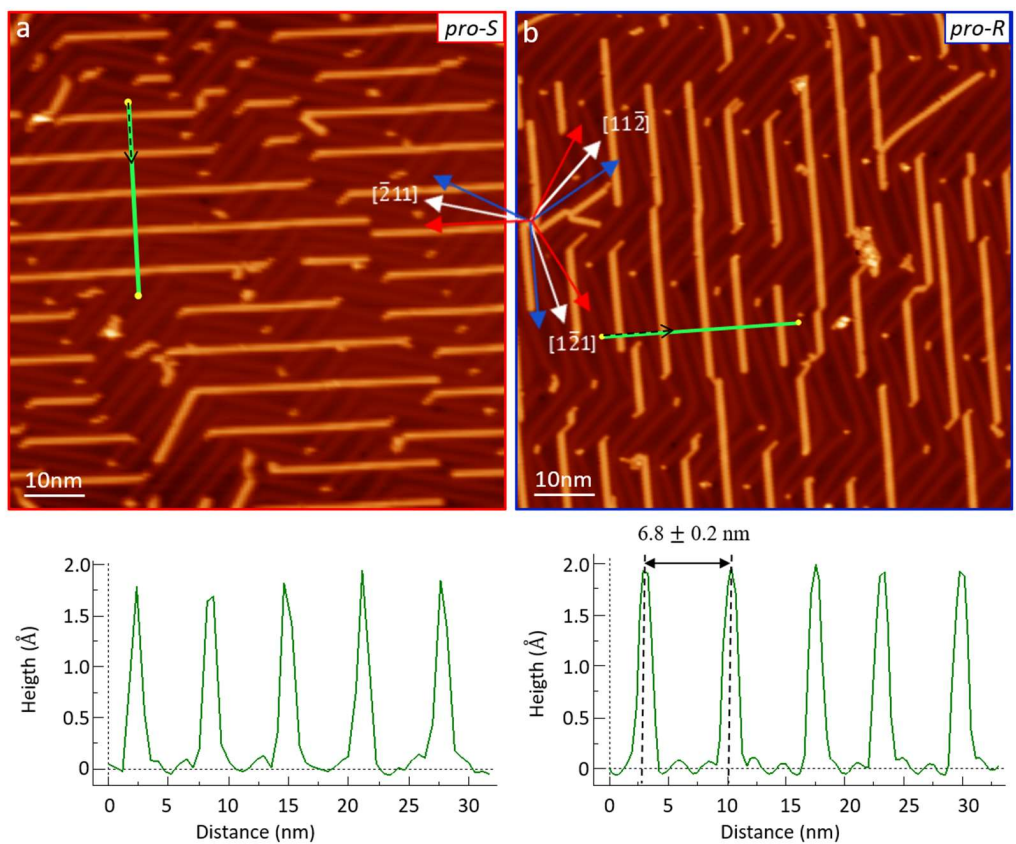

**Fig. S8.** STM images ($V_s = 1.0V$, $I_t = 200pA$; $V_s = 0.5V$, $I_t = 150pA$) showing the well-defined preferred interspacing observed for both (a) (*pro-S*)-GNRs and (b) (*pro-R*)-GNRs. Green lines correspond to the height profiles at the bottom of each image. Dashed black lines indicate the height profile directions.